\documentclass[12pt,preprint]{aastex} 

\slugcomment{Astrophysical Journal, in press}

\shorttitle{Solar Minimum X-ray emission}
\shortauthors{Sylwester et al.}

\begin{document}

\title{SPHINX MEASUREMENTS OF THE 2009 SOLAR MINIMUM X-RAY EMISSION }


\author{J. SYLWESTER\altaffilmark{1}, M. KOWALINSKI\altaffilmark{1}, S. GBUREK\altaffilmark{1},  M. SIARKOWSKI\altaffilmark{1}, S. KUZIN\altaffilmark{2}, \\ F. FARNIK\altaffilmark{3},
 F. REALE\altaffilmark{4},  K. J. H. PHILLIPS\altaffilmark{5},  J. BAKA{\L}A\altaffilmark{1}, M. GRYCIUK\altaffilmark{1}, \\
 P. PODGORSKI\altaffilmark{1}, AND B. SYLWESTER\altaffilmark{1}}

\altaffiltext{1}{Space Research Centre, Polish Academy of Sciences, 51-622, Kopernika~11, Wroc{\l}aw, Poland; \email{js@cbk.pan.wroc.pl}}

\altaffiltext{2}{P. N. Lebedev Physical Institute (FIAN), Russian Academy of Sciences, Leninsky Prospect 53, Moscow 119991, Russia}

\altaffiltext{3}{Astronomical Institute, Ond\v{r}ejov Observatory, Czech Republic}

\altaffiltext{4}{Dipartimento di Fisica, Universit\'{a} di Palermo, Palermo, Italy, and INAF, Osservatorio Astronomico di Palermo, Palermo, Italy}

\altaffiltext{5}{Mullard Space Science Laboratory, University College London, Holmbury St Mary, Dorking,
Surrey RH5 6NT, UK }

\begin{abstract}
The SphinX X-ray spectrophotometer on the {\em CORONAS-PHOTON} spacecraft measured soft X-ray emission in the 1--15~keV energy range during the deep solar minimum of 2009 with a sensitivity much greater than {\em GOES}. Several intervals are identified when the X-ray flux was exceptionally low, and the flux and solar X-ray luminosity are estimated. Spectral fits to the emission at these times give temperatures of 1.7--1.9~MK and emission measures between $4\times 10^{47}$~cm$^{-3}$ and $1.1\times 10^{48}$~cm$^{-3}$. Comparing SphinX emission with that from the {\em Hinode} X-ray Telescope, we deduce that most of the emission is from general coronal structures rather than confined features like bright points. For one of 27 intervals of exceptionally low activity identified in the SphinX data, the Sun's X-ray luminosity in an energy range roughly extrapolated to that of {\em ROSAT} (0.1--2.4~keV) was less than most nearby K and M dwarfs.
\end{abstract}


\section{INTRODUCTION}\label{intro}

The period of low solar activity during most of 2009 has been reported on extensively, though relatively little attention thus far has been paid to the amount of soft X-ray emission which is an important guide to the state of the solar corona's magnetic complexity. For much of 2009, the emission in the {\it Geostationary Operational Environmental Satellites} ({\em GOES}) longer-wavelength channel (range 1--8~\AA) was below its threshold ($3.7 \times 10^{-9}$ W~m$^{-2}$). Here we report on results from a highly sensitive soft X-ray spectrophotometer SphinX (Solar PHotometer IN X-rays), part of the TESIS package which was flown on board the Russian {\em CORONAS-PHOTON} spacecraft. Unlike {\em GOES}, SphinX successfully recorded soft X-ray emission over this entire period. The SphinX energy range was 1--15~keV, comparable to that of the {\em GOES} longer-wavelength channel. {\em CORONAS-PHOTON}, which was part of the international {\em Living With a Star} program, was launched on 2009 January~30 into a near-polar orbit, and operated till 2009 December when spacecraft power failures terminated the mission. SphinX measurements were made from 2009 February~20 to November~28.

Results from SphinX, preliminary versions of which have already appeared \citep{syl11b,gbu11a,gbu11b}, indicate that on a number of periods during 2009 the 1--8~\AA\ emission was some 20 times less than the {\em GOES} A1 level (equal to a flux of $10^{-8}$ W~m$^{-2}$), far less than the lowest level in the previous (1998) solar minimum ($5.2 \times 10^{-9}$~W~m$^{-2}$). The spectral capabilities of SphinX allow the temperature and emission measure to be obtained for these periods, with isothermal assumption for the emission from the whole corona. From these values, a rough extrapolation to the 0.1--2.4~keV energy range is made so that estimates can be made of the solar X-ray luminosity in a band comparable to the {\em Student Nitric Oxide Explorer} ({\em SNOE}) in 1998--1999 and to that measured by the Position Sensitive Proportional Counter (PSPC) instrument on the {\em ROSAT} spacecraft. These comparisons suggest that the Sun was a very weak X-ray emitter in 2009 compared with the time of {\em SNOE} and with nearby stars K and M stars surveyed by {\em ROSAT}.

\section{THE SPHINX INSTRUMENT AND PERFORMANCE}

The SphinX instrument \citep{syl08,gbu11a,gbu11b} was constructed at the \/\/Space Research Centre, Polish Academy of Sciences, in Wroc{\l}aw, Poland. Three Peltier-cooled PIN diodes (D1, D2, and D3) viewed the Sun behind precisely measured apertures with areas 21.50, 0.495, and 0.0101~mm$^2$ respectively. The detectors, manufactured by Amptek (Bedford, MA), are silicon wafers 500~$\mu$m thick with 12.5~$\mu$m beryllium windows, and were sensitive to $\sim 1.2-15$~keV (0.8--10~\AA) X-rays. The instrument was intensity-calibrated to better than 5\% accuracy using various characteristic X-ray lines viewed through a 35-m-long pipe at the X-ray Astronomy Calibration and Testing (XACT) facility (Palermo, Italy) and at the BESSY II (Berlin, Germany) synchrotron source. The D1 detector, which had a sensitivity approximately 100 times greater than the {\em GOES} 1--8~\AA\ channel, was designed to measure very low X-ray fluxes. For moderate levels of X-ray emission, this detector was liable to pulse pile-up and saturation effects. Detectors D2 and D3 were intended for measuring the flux at higher levels of X-ray emission, though throughout the mission detector D3 recorded only noise while measurements from D2 had a mostly poor signal-to-noise ratio. A fourth detector, designed to measure X-ray fluorescence within the instrument excited by solar radiation, was not turned on during the mission owing to the very low level of solar flux. The detector energy ranges from pre-launch calibration measurements were 1.2--14.9~keV (D1) and 1.0--14.9~keV (D2). Measured values of spectral resolution (FWHM) were 464~eV (D1) and 319~eV (D2). They are an improvement over those of the {\em RHESSI} spectrometer detectors ($\sim 1$~keV), and comparable to that  of the X-ray Spectrometer solid state detectors -- 600~eV at 5.9~keV -- on both {\em MErcury Surface, Space Environment, GEochemistry and Ranging} ({\em MESSENGER}: \cite{sch07}) and  {\em Near Earth Asteroid Rendezvous} ({\em NEAR}: \cite{gol97}). At the resolution of SphinX, some line features (composed of unresolved groups of lines) are recognizable, including the Fe line feature at 6.7~keV during small flares that occasionally occurred in 2009. SphinX spectra were registered both in 256 channels and in ancillary four-channel spectra (so-called basic mode) during the mission. In the basic mode, total photon counts were recorded in the ranges 1.5--3.0~keV and 3.0--14.9~keV (D1) and 1.0--3.0~keV and 3.0--14.9~keV (D2), allowing X-ray light curves to be plotted, while another channel recorded particle emission.

The plane of the near-polar orbit (inclination $82.5^\circ$) of {\em CORONAS-PHOTON} was almost perpendicular to the Sun's direction, allowing the instruments including SphinX to view the Sun for at least 60~minutes of the 96-minute orbital period, with full-Sun coverage for two-week periods in April and July. However, there were interruptions to the observations due to enhanced particle background for up to 4 passages per orbit through the auroral ovals and  two or three passages through the South Atlantic Anomaly in a 24-hour period. Solar measurements were possible for 51\% of the time that SphinX was turned on, or 37.5\% of the total time that {\em CORONAS-PHOTON} operated.

An extensive grid of model spectra calculated with the {\sc chianti} atomic database and code \citep{der97,der09} with various element abundance sets and ionization fractions  was calculated to assist in interpretation of SphinX spectra. Comparison with observed spectra allows electron temperature ($T_e$) and volume emission measure $EM = N_e^2 V$ ($N_e = $ electron density, $V = $ emitting volume) to be deduced from the observed emission on an isothermal assumption.

An anomaly occurred in the D1 detector on April~6 when enhanced count rates, unassociated with solar activity, were observed over a two-hour period. Within about four hours after this, according to a pre-planned sequence, new command software was uploaded to the spacecraft, resulting in a nearly full recovery. The anomaly therefore had minimal effects and did not affect the data analysis discussed here.

\section{X-RAY EMISSION FROM SPHINX AND EMISSION MEASURE ANALYSIS}

\begin{figure}
\begin{center}
\epsscale{1.4}
\includegraphics[width=12cm,angle=0]{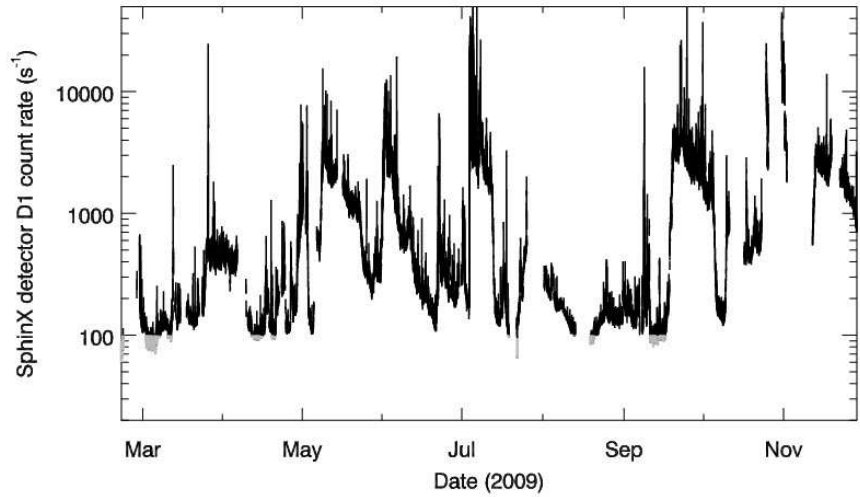}
\caption{The X-ray emission as recorded by the SphinX D1 detector (in counts s$^{-1}$), from daily averages over the period 2009 February 20 -- November 28. Portions of the curve marked gray (blue in the on-line journal) show periods selected for especially low activity. The detector count rates are as observed, i.e. not corrected for the slight variation in the Earth--Sun distance.  (A color version of this figure is available in the online journal.) } \label{D1_countrate_blue}
\end{center}
\end{figure}

Either the D1 or D2 detectors may be used to monitor the coronal X-ray activity in 2009. Since the saturation and pile-up effects with the D1 detector are apparent only for flares, this detector is the preferred one for tracking the general, non-flare X-ray emission. Figure~\ref{D1_countrate_blue} shows the D1 light curve (in photon counts s$^{-1}$) over the period 2009 February 20 to November 28, constructed from intervals free from obvious particle events. The passage of active regions is evident from the increase of the D1 count rate in periods of up to about 14 days, the length of time for a long-lasting active region to rotate across the Sun's visible hemisphere. The largest peak occurred in early July, when NOAA active region 11024 dominated the emission; this period has already been discussed by \cite{eng11} and \cite{syl11b}. Apart from flare events, {\em GOES} did not record any emission above its detection limit over this entire period. Times of particularly low emission are indicated in Figure~\ref{D1_countrate_blue}. Twenty-seven such intervals were identified, each of several hours' duration, the time ranges of which are given in Table~\ref{low_flux_intervals}. No major active regions were recorded at these times, which may be regarded as among the lowest-activity periods during the 2009 solar minimum.

\begin{deluxetable}{rlccccc}
\tabletypesize{\scriptsize} \tablecaption{S{\sc phinX} M{\sc easurements} {\sc during} I{\sc ntervals of very low} A{\sc ctivity in } 2009 \label{low_flux_intervals}} \tablewidth{0pt}

\tablehead{\colhead{Int.} & \colhead{Time range (dates are 2009)} & \colhead{$T_e$ (MK)$^a$} & \colhead{log$_{10}$~EM$^a$} & \colhead{SphinX X-ray (1--8 \AA) flux}  & \colhead{1--15 keV luminosity} \\
&&&\colhead{(EM in cm$^{-3}$)}& \colhead{(W m$^{-2}$)$^b$} & \colhead{(erg s$^{-1}$)$^b$} \\ }

\startdata

1  & Feb 20 20:12 -- Feb 21 05:46  & 1.73 (.04) & 47.81 (.01) & 4.91 (-10) & 4.63 (22) \\
2  & Feb 21 06:33 -- Feb 21 15:21  & 1.81 (.04) & 47.72 (.01) & 6.62 (-10) & 5.10 (22) \\
3  & Mar 2 12:23 -- Mar 2 17:57   & 1.89 (.04)  & 47.61 (.01) & 7.70 (-10) & 5.15 (22) \\
4  & Mar 3 06:02 -- Mar 4 00:15 & 1.88 (.05)  & 47.62 (.01)   & 7.60 (-10) & 5.15 (22) \\
5  & Mar 4 09:05 -- Mar 4 17:49 & 1.84 (.04)  & 47.68 (.01)   & 7.21 (-10) & 5.22 (22) \\
6  & Mar 4 23:21 -- Mar 5 11:21 & 1.79 (.03) & 47.78 (.01)    & 6.48 (-10) & 5.33 (22) \\
7  & Mar 5 18:42 -- Mar 5 20:56 & 1.80 (.05) & 47.69 (.01)    & 5.90 (-10) & 4.66 (22) \\
8  & Mar 11 15:02 -- Mar 11 20:34 & 1.79 (.04) & 47.83 (.01)  & 7.35 (-10) & 6.03 (22) \\
9  & Apr 12 13:05 -- Apr 12 17:30 & 1.74 (.03) & 47.93 (.01)  & 6.68 (-10) & 6.16 (22) \\
10 & Apr 13 15:50 -- Apr 13 20:12 & 1.72 (.02) & 47.95 (.01) & 6.38 (-10) & 6.11 (22) \\
11 & Apr 14 05:33 -- Apr 14 11:19 & 1.74 (.03) & 47.92 (.01) & 6.66 (-10) & 6.10 (22) \\
12 & Apr 19 10:31 -- Apr 19 14:22 & 1.85 (.03) & 47.75 (.01) & 8.64 (-10) & 6.16 (22) \\
13 & Apr 19 20:11 -- Apr 20 05:19 & 1.92 (.03) & 47.64 (.01) & 9.85 (-10) & 6.20 (22) \\
14 & Apr 20 06:09 -- Apr 20 12:59 & 1.88 (.02) & 47.73 (.01) & 9.76 (-10) & 6.56 (22) \\
15 & Jul 18 14:13 -- Jul 18 16:54 & 1.72 (.03) & 48.01 (.01) & 7.28 (-10) & 7.01 (22) \\
16 & Jul 18 18:30 -- Jul 19 02:09 & 1.73 (.03) & 48.00 (.01) & 7.59 (-10) & 7.09 (22) \\
17 & Jul 22 00:52 -- Jul 22 04:13 & 1.75 (.03) & 47.97 (.01) & 8.17 (-10) & 7.31 (22) \\
18 & Aug 12 19:14 -- Aug 13 09:36 & 1.74 (.03) & 47.98 (.01) & 7.79 (-10) & 7.11 (22) \\
19 & Aug 19 05:39 -- Aug 19 20:01 & 1.72 (.03) & 47.96 (.01) & 6.34 (-10) & 6.14 (22) \\
20 & Aug 19 23:11 -- Aug 20 10:43 & 1.68 (.03) & 48.05 (.01) & 5.99 (-10) & 6.44 (22) \\
21 & Aug 20 12:00 -- Aug 20 18:40 & 1.87 (.04) & 47.75 (.01) & 9.89 (-10) & 6.81 (22) \\
22 & Sep 11 09:30 -- Sep 11 22:12 & 1.79 (.04) & 47.83 (.01) & 7.64 (-10) & 6.16 (22) \\
23 & Sep 11 22:54 -- Sep 12 12:37 & 1.70 (.03) & 47.99 (.01) & 6.05 (-10) & 6.12 (22) \\
24 & Sep 13 02:17 -- Sep 13 07:46 & 1.69 (.03) & 48.01 (.01) & 6.10 (-10) & 6.27 (22) \\
25 & Sep 13 19:33 -- Sep 13 21:11 & 1.68 (.06) & 48.02 (.02) & 5.80 (-10) & 6.13 (22) \\
26 & Sep 14 21:04 -- Sep 15 07:37 & 1.79 (.03) & 47.84 (.01) & 7.79 (-10) & 6.28 (22) \\
27 & Sep 16 01:50 -- Sep 16 07:34 & 1.71 (.02) & 47.99 (.09) & 6.53 (-10) & 6.48 (22) \\
\\

\enddata

\tablenotetext{a}{Temperatures and logarithms of emission measures are from spectral fits using {\sc chianti} (v.~6.1), with uncertainties in parentheses. X-ray fluxes (1--8~\AA, col.~5) are estimated from these temperatures and emission measures, as are the 1--15~keV luminosities (col.~6). Emission measures, X-ray fluxes and luminosities are corrected for 1 AU.} \tablenotetext{b}{Numbers in parentheses indicate powers of 10.}
\end{deluxetable}

Using spectral data from SphinX, averaged X-ray spectra during each of the twenty-seven intervals were analyzed using pre-launch measurements of the instrument's  sensitivity and spectral resolution. These averaged spectra were fitted with model {\sc chianti} (version~6.1: \cite{der09}) spectra. Ionization fractions from \cite{bry09} were used for these model spectra. A coronal element abundance set \citep{fel92} was chosen, these differing from photospheric abundances (e.g. \cite{asp09}) for elements with low ($\lesssim 10$~eV) first ionization potentials (FIPs), being enhanced over photospheric abundances by factors of up to four. Some recent results \citep{syl10a,syl10b,syl11a,phi12} suggest that the enhancement factors are different from four for some elements, though here we neglect the effect on the total emission in the 1.2--12~keV range. Table~\ref{low_flux_intervals} gives temperatures and logarithms of volume emission measures obtained from the {\sc chianti} spectral fits. From these, total X-ray fluxes in the 1--8~\AA\ range are estimated as are total X-ray luminosities (1--15~keV band), assuming that the Sun's X-ray emission on the hemisphere visible from the {\em CORONAS-PHOTON} spacecraft is typical of the whole Sun. Emission measures, X-ray fluxes, and luminosities are corrected for an Earth--Sun distance of 1~AU.

Estimates of temperature and volume emission measure from X-ray spectra are related in an inverse manner, since an overestimate of temperature tends to lead to an underestimate of emission measure for a given spectrum and vice versa. This is evident if the temperatures and emission measures from SphinX spectra in Table~\ref{low_flux_intervals} are examined or plotted against each other. Thus, this apparent inverse relation is not a physical one and in any case is opposite to that expected from simple loop models such as those of \cite{ros78} or observational results for X-ray flares using {\em GOES} emission ratios \citep{fel96}. A more meaningful result is the mean of the various parameters in the twenty-seven intervals listed in Table~\ref{low_flux_intervals}: temperature $1.78 \pm 0.07$~MK, logarithm of emission measure (in cm$^{-3}$): $47.85 \pm 0.14$; 1--8~\AA\ flux ($10^{-10}$ W~m$^{-2}$): $7.21 \pm 1.26$; 1--15~keV X-ray luminosity ($10^{22}$ erg s$^{-1}$): $6.07 \pm 0.73$.

\begin{figure}
\epsscale{1.1}
\includegraphics[width=12cm,angle=0]{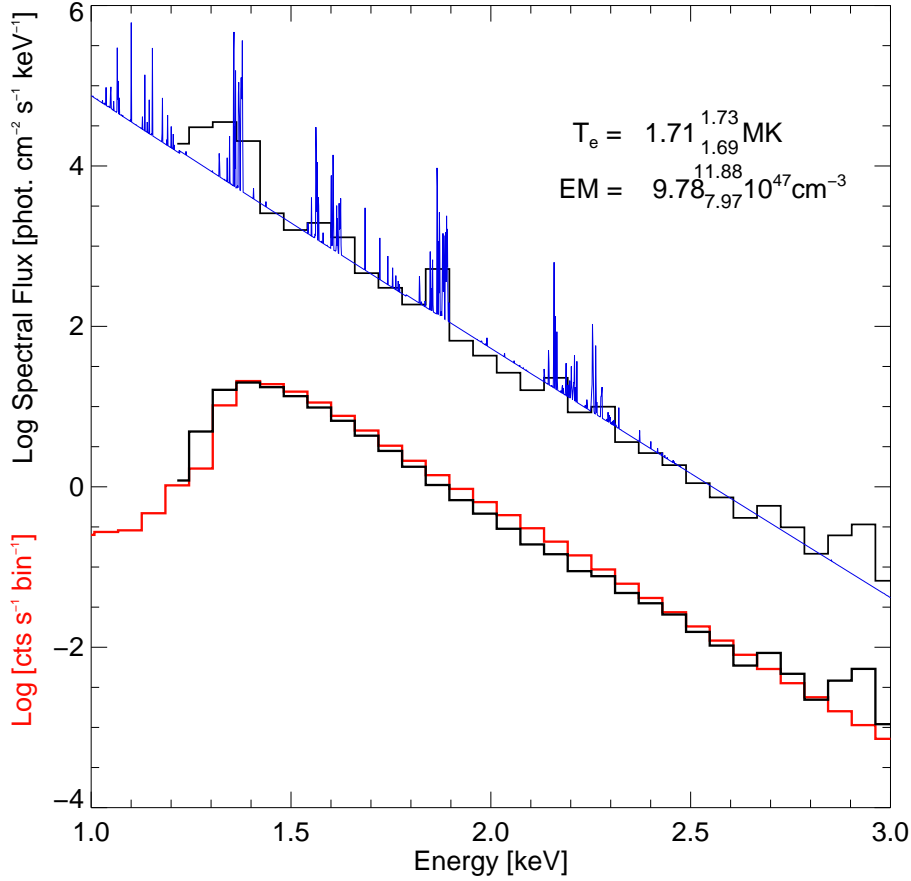}
\caption{ Averaged photon spectrum in the 1--3~keV range (upper histogram) over a time period on 2009 September~16 between 01:50~UT and 07:33~UT, made up of intervals when the total SphinX D1 count rate was below 110 counts s$^{-1}$. The energy bins correspond to those in the count rate spectrum (lower histograms). The blue curve is the {\sc chianti} photon spectrum at a few eV resolution showing principal line groups. In the count rate spectra, the black histogram is the observed SphinX spectrum, and the red histogram shows the best fit to the count rate spectrum with estimated temperature and emission measure  indicated in the legend. (A color version of this figure is available in the online journal.) } \label{aver_sp}
\end{figure}

Figure~\ref{aver_sp} shows an averaged SphinX count rate spectrum (histograms in the lower part) and photon spectrum (upper part) over the time interval 27 in Table~\ref{low_flux_intervals}, 2009 September~16 (01:50--07:33~UT), a range of 5.7 hours (20580~s). A total of $\sim 10^7$ counts were accumulated in this spectrum.  An isothermal spectral fit to the count rate spectrum is shown, based on {\sc chianti} theoretical spectra with coronal abundances \citep{fel92}, a temperature of $1.71\pm 0.02$~MK and emission measure $(9.78\pm 2.0) \times 10^{47}$~cm$^{-3}$. Line features making a strong contribution include those at $\sim 1.3$~keV and $\sim 1.6$~keV, due to \ion{Mg}{11} lines, and those at 1.85~keV and 2.1~keV due to groups of \ion{Si}{12} dielectronic satellites. As Mg and Si are low-FIP elements, the fluxes of these line groups depend on the choice of coronal or photospheric abundance sets in the calculated spectrum. The continuum in this range is mostly made up of free--bound emission, with chief contributing elements He, C, and O. As these are high-FIP elements, the continuum emission is not so sensitive to the choice of element abundance set. For the photon spectrum shown, the spectral fluxes (photon cm$^{-2}$~s$^{-1}$~keV$^{-1}$) were derived from inversion of the full detector response matrix. The {\sc chianti} photon spectrum with a few eV resolution is also shown to indicate the location of the principal groups of spectral lines.

\begin{figure}
\epsscale{1.1}
\plottwo{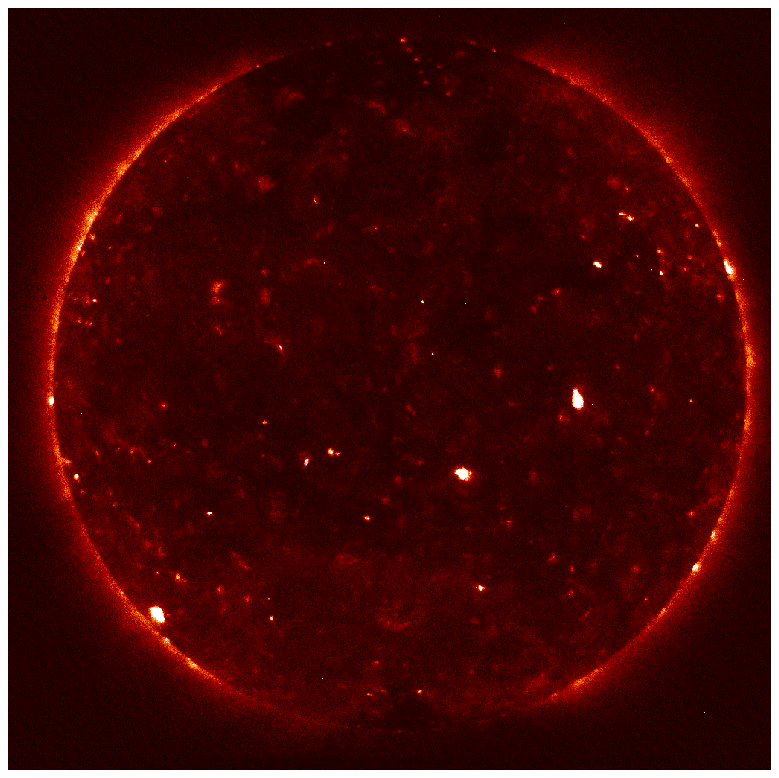}{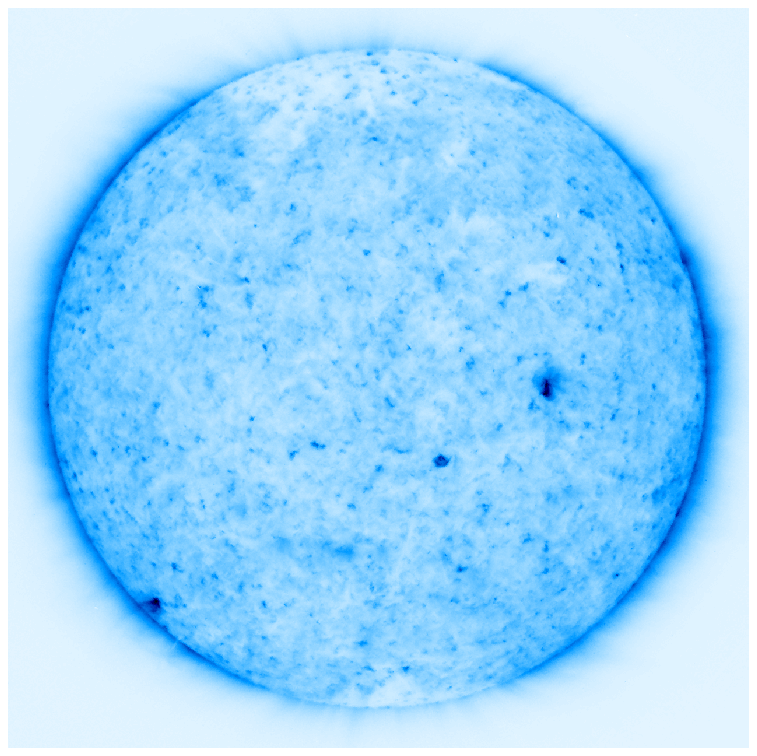}
\caption{(Left) {\em Hinode} XRT image (Ti-poly filter) taken on 2009 September~15 at 15:47:31~UT, between intervals 26 and 27 of the low-flux levels recorded by SphinX (Table~\ref{low_flux_intervals}). Despite the impression from this figure, the general corona dominates the total emission in this image, the X-ray bright points accounting for only a small fraction. (Right) TESIS image in its 171~\AA\ filter taken on September~15 (16:24:27~UT). Solar north is towards the top of both images. (A color version of this figure is available in the on-line journal.) }  \label{XRT_Ti_poly_image}
\end{figure}

Estimates of temperature and emission measure from the isothermal fit to SphinX spectra during low levels of X-ray emission in 2009 can be combined with a series of XRT images on September~15 to find where the bulk of the visible-hemisphere X-ray coronal emission is formed. Figure~\ref{XRT_Ti_poly_image} (left panel) is the full-Sun XRT image taken in its Ti-poly filter at 15:47~UT. This filter has a nominal passband (one-tenth power points) equal to 5.4--16.6~\AA\ (0.7--2.3~keV), similar to the range of detectable emission in SphinX (Figure~\ref{aver_sp}). The XRT image is characterized by general coronal emission and several bright points with measured fluxes of up to 740 data numbers (DN) s$^{-1}$. The three most intense of the bright points, with fluxes above 50~DN~s$^{-1}$, account for only 1.6\% of the total X-ray emission. The general coronal emission is not uniform, with coronal holes evident at particularly the south pole and at low latitudes. About 58\% of the visible hemisphere is covered by emission equal to between 3~DN~s$^{-1}$ and 10~DN~s$^{-1}$ which we identify with the general corona. In the TESIS 171~\AA\ coronal image (right panel) taken shortly after (16:24~UT), the emission features including the bright points and coronal holes are similar to those in the XRT Ti-poly image, though with the addition of polar plumes which are clearly present at the south pole. The emission in the TESIS image is primarily from \ion{Fe}{9} and \ion{Fe}{10} lines emitted at $\sim 1$~MK, similar to the estimated temperature from the SphinX spectrum of Figure~\ref{aver_sp}.

The volume of the corona on the visible hemisphere, estimated from the proportion covering the visible hemisphere and X-ray intensity fall-off at the limb, is $4.2 \times 10^{31}$~cm$^3$, which, combined with the SphinX emission measure $N_e^2 V$ for interval~27 in Table~\ref{low_flux_intervals}, gives the electron density $N_e$ averaged over the corona to be $1.5 \times 10^8$~cm$^{-3}$. This approximate calculation, though it neglects the likely presence of fine structure, is consistent with estimates from visible coronal measurements (\cite{leb73}: $(1-2)\times 10^8$~cm$^{-3}$), measurements from the {\em CORONAS-F} SPIRIT spectroheliograph (\cite{she09}: $(2.5-6)\times 10^8$~cm$^{-3}$), and coronal hole densities from the {\em SOHO} SUMER instrument (\cite{dos97}: $\sim 10^8$~cm$^{-3}$).

\section{SOLAR X-RAY LUMINOSITY}

The solar X-ray luminosity in the SphinX detector~D1 energy range (1.2--14.9~keV) may be simply obtained from X-ray fluxes, corrected for 1~AU, by the area of a sphere with radius 1~AU ($2.81\times 10^{27}$~cm$^2$), assuming that the emission from the visible hemisphere typifies the entire Sun. These are given in Table~\ref{low_flux_intervals}; the values range from $4.6 \times 10^{22}$ to $7.3 \times 10^{22}$~erg~s$^{-1}$. SphinX was the only instrument in operation monitoring solar radiation in the 1--15~keV or comparable range during the 2009 solar minimum, while there was no equivalent instrument during the solar minimum of 1996. The Soft X-ray Telescope (SXT) on {\em Yohkoh}, operational from 1991 to 2001, had a variety of transmission filters including a beryllium filter with an energy range near that of SphinX, but full-Sun images were not often taken in this filter. An analysis by \cite{act96} of thin aluminum filter full-Sun images in the softer $\sim 0.3-3$~keV range over the period 1991 November to 1995 September indicates a minimum X-ray luminosity of $2.5 \times 10^{25}$~erg~s$^{-1}$ in 1995 July, several months before the solar minimum between Cycles 22 and 23. This value is much larger than that obtained from SphinX detector~D1 owing to the extra emission in the 0.3--1.2~keV range that the SXT thin aluminum filter is sensitive to.

Measurements of the solar X-ray emission have previously been made during other times of low activity. Those of the ({\em SNOE}) spacecraft (operational 1998--2003: \cite{jud03}) cover the range 0.1--2.4~keV, and so included much softer energies (0.1--1.2~keV) than those available to SphinX detector D1. This interval, similar to that of {\em ROSAT} PSPC instrument used for the {\em ROSAT} All-Sky Survey (RASS, \cite{sch95}), includes H-like and He-like C, N, and O resonance lines and strong lines of Mg, Si, and Fe emitted at coronal temperatures as well as a few lines emitted at lower temperatures characteristic of the solar transition region. The {\sc chianti} isothermal fit to the SphinX spectrum shown in Figure~\ref{aver_sp} can be used to give a crude estimate of the X-ray luminosity in the 0.1--2.4~keV energy range for comparison. With $T = 1.71$~MK and $EM = 9.8 \times 10^{47}$~cm$^{-3}$ and using {\sc chianti}, the 0.1--2.4-keV X-ray luminosity, or $L_{RASS}$, is calculated to be log~$L_{RASS} = 25.2$ ($L_{RASS}$ in erg~s$^{-1}$). This estimate compares with those of \cite{jud03} from {\em SNOE}, which range from log~$L_{RASS} = 27.1$ to 27.75 in the 1998--1999 period: the 1996--1997 minimum was therefore missed. However, extrapolation by \cite{jud03} to the minimum indicates log~$L_{RASS}$ might have been as low as 26.8. The SphinX data thus indicates that the 2009 minimum was much deeper in soft X-ray emission. The survey of 144 K and M dwarf stars within 7~parsecs surveyed by {\em ROSAT} PSPC indicates that only 15\% of the stars have comparable or lower luminosities. 

A differential emission measure (DEM) analysis might improve this estimate by taking into account the softer X-ray emission, and is in principle possible from the discrimination available in the several broad-band filters of the {\em Hinode} XRT instrument. However, the variety of DEM solutions obtained from a sequence of XRT full-Sun images near in time to the spectrum shown in Figure~\ref{aver_sp} resulted in X-ray spectra that were far too flat compared with that observed: thus, at energies of $\sim 3$~keV, there were several orders of magnitude more emission. It would appear that the limitations of DEM analysis in the X-ray range under discussions are too severe for meaningful analysis.

\section{SUMMARY}

Measurements with the SphinX spectrophotometer in the 1--15~keV energy range during the deep solar minimum of 2009 have shown the low level of soft X-ray emission in this period to a much improved extent than is available through {\em GOES} which was below its threshold throughout the period apart from occasional flares. Periods of exceptionally low activity have been identified, with spectra showing significant flux only in the 1--3~keV range and indicating that general coronal temperatures were between 1.7~MK and 1.9~MK, with emission measures between about $4 \times 10^{47}$~cm$^{-3}$ and $1.1 \times 10^{48}$~cm$^{-3}$. Combined with X-ray imaging data from the {\em Hinode} XRT, a lower limit for the coronal density is $\sim 1.5 \times 10^8$~cm$^{-3}$. The luminosities in the SphinX energy range study have been extrapolated to the {\em SNOE} and {\em ROSAT} 0.1--2.4~keV range using an isothermal fit to the low-activity spectrum on September~16. Though very rough, these estimates indicate that on September~16 and at dates with similar X-ray activity levels the Sun's X-ray luminosity from SphinX was much lower than in the 1996--1997 minimum and places the Sun in the lowest-activity 15\% of nearby K and M dwarf stars. This study should be of interest in view of the current predictions for the low level of the next solar minimum \citep{alt11}, and shows the need for instrumentation similar to SphinX if, as in the 2009 solar minimum,  {\em GOES}  lacks the sensitivity to measure X-ray fluxes on a continuous basis. Space-borne instruments such as the next-generation SphinX (SPHINX-NG) on a forthcoming Polish-US nanosatellite (``CubeSat") spacecraft and STIX on {\em Solar Orbiter} should be in a position to estimate future solar X-ray levels in the event of forthcoming deep solar minima.

\acknowledgments

We acknowledge financial support from the European Commission's Seventh Framework Programme (FP7/2007--2013) under grant agreement No. 218816 (SOTERIA project, www.soteria-space.eu), the Polish Ministry of Education and Science Grant N N203 381736,  and the UK--Royal Society/Polish Academy of Sciences International Joint Project (grant number 2006/R3) for travel support. {\sc chianti} is a collaborative project involving George Mason University, the University of Michigan (USA) and the University of Cambridge (UK).  We thank K. Reeves and M. Weber for considerable help with interpretation of {\it Hinode} XRT data.

\end{document}